# On detection of narrow angle e+e- pairs from dark photon decays

S. N. Gninenko[1], A.V. Dermenev[1], S.V. Donskov[2], S.B. Kuleshov[5], V.A. Matveev[1,3], V.V. Myalkovskiy[3], V.D. Peshekhonov[3], V.A. Poliakov[2], A. A. Savenkov[3], V.O. Tikhomirov[4,6], I.A. Zhukov[3]

[1] Institute for Nuclear Research, 117 312 Moscow, Russia
[2] State Research Center of the Russian Federation, Institute for High Energy Physics, 142281 Protvino, Russia
[3] Joint Institute for Nuclear Research, , 141 280 Dubna, Russia
[4] P.N. Lebedev Physics Institute, Moscow, Russia, 119 991 Moscow, Russia;
[5] Universidad Técnica Federico Santa María, 2390123 Valparaiso, Chile
[6] National Research Nuclear University MEPhI, 115409 Moscow, Russia

**Abstract**

A class of models of dark sectors consider new very weak interaction between the ordinary and dark matter transmitted by U'(1) gauge bosons A' (dark photons) mixing with our photons. If such A's exist, they could be searched for in a light-shining-through-a-wall experiment with a high energy electron beam from the CERN SPS. The proposed search scheme suggests detection of the e+e- pairs produced in the A' -> e+e- decay with a very small opening angle. Coordinate chambers based on the thin-wall drift tubes with a minimal material budget and a two-hit resolution for e+ and e- tracks separated by more than 0.5 mm are considered as an option for detecting such pairs.

## 1. INTRODUCTION

New ambitious research projects in the field of particle physics which are discussed nowadays will naturally require not only optimization of potentialities of acceleration facilities but also a development of particle detection technique by means of modifying existing detectors, data acquisition systems and data analysis procedures. Therefore, it is important to investigate the new capabilities of the well-known detectors as well as to develop detectors based on completely new principles.

The origin and properties of dark matter is one of the main problems of the modern cosmology and particle physics. Some dark matter models assume the existence of dark sector which contains particles described by singlet fields with respect to the gauge group $SU(3)_C \times SU(2)_L \times U(1)_Y$ of the Standard Model. In a class of these models, in addition to gravity, a very weak interaction between normal and dark matter can exist. It is assumed that the interaction can be transmitted by massive $U'(1)$ gauge bosons (dark photons A') which could mix with ordinary photons. The dark photons, A', could have mass in sub-GeV range and coupling to photons lying in the experimentally accessible and theoretically motivated parameter space. The topic of the search for dark photons is in the research program of several laboratories, see e.g. [1], and references therein.

The proposal P348 for the search for A's at the CERN SPS relies on detecting e+e- pairs with a small opening angle [2]. A feasible coordinate detector can be constructed on the base of the thin-walled drift tubes (straw tubes) which are characterized by the smallest material budget compared to other types particle detectors. However, the straw tube detectors should provide the two-hit resolution similar to other gas-filled detectors. For example, the two-hit resolution of a GEM detector can reach 0.5 mm [3]. In order to investigate the feasibility of employing the coordinate detectors based on straw tubes for the setup suggested in the proposal P348 for registering e+e- pairs, the high precision straw tubes have been constructed and studied at JINR.

## 2. Beam Tests and the Experimental Setup

The observation of A' decay presents a challenge for the detector design and performance. The experimental setup specifically designed to search for the A′ -> e+e− decays [2] is schematically shown in Fig. 1. The experiment will employ the CERN SPS H4 e− beam, which is produced at the target T2 of the CERN SPS and transported to the detector in an evacuated beamline with a momentum from 10 up to 300 GeV/*c*. The typical maximal beam intensity at ~ 30--100 GeV/*c* is of the order of $5 \times 10^6$ e− per SPS spill of 4.8 s with $10^{12}$ protons on the target.

The setup shown in Fig.1 is equipped with a high density electromagnetic calorimeter ECAL1 to detect e− primary interactions, high efficiency veto counters V1 and V2, two thin-wall straw tube chambers ST1, ST2, an electromagnetic calorimeter ECAL2 located at the downstream end of the A′ decay volume DV to detect e+e− pairs from A′ -> e+e− decays in flight, and a hadronic calorimeter (HCAL) used mainly for the search of A′ -> invisible decay mode.

The A′s are produced through the mixing with bremsstrahlung photons from the electron scattering off nuclei in the ECAL1, e- + Z --> e'- + Z + A', as shown in Fig. 1. This reaction typically occurs in the detector at depth of a few radiation lengths. The bremsstrahlung A' then penetrates the rest of the calorimeter ECAL1 and the veto counter V1 without interactions, and decays in flight into the e+e− pair in the decay volume DV of the length $L$. The signal events would have a unique signature characterized by two electromagnetic showers which would develop separately along the beam axis, as shown in Fig. 1. The first shower is produced in the target, the electromagnetic calorimeter ECAL1, by the scattered electron, e'-, and the second one is observed in the electromagnetic calorimeter ECAL2 as a total signal caused by the energy deposition from the e+e- pair with small opening angle further called "narrow decaying pair". The tracks produced by the e+e- pair are registered in the straw chambers ST1 and ST2, while the energy of the e+e- pair is measures by the electromagnetic calorimeter ECAL2.

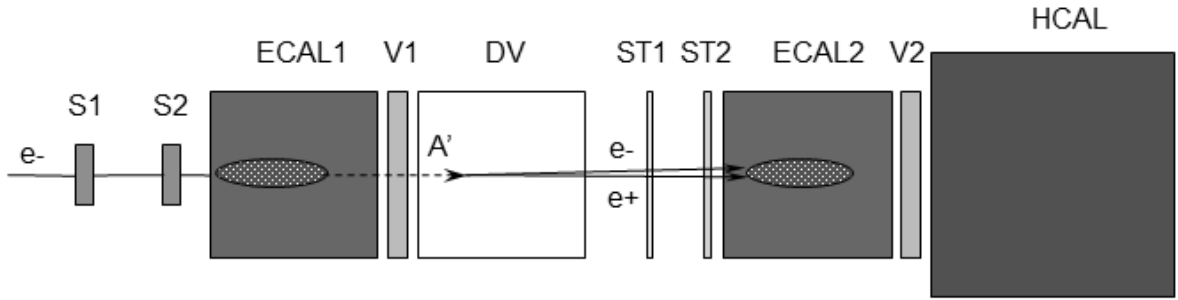

FIG. 1: Schematic illustration of the setup to search for dark photons in a light-shining-through-a-wall type experiment at high energies. The primary electron beam is defined by two scintillation counters S1 and S2. The energy deposited by the incident electron in the ECAL1 is accompanied by the emission of bremsstrahlung A' in the reaction eZ -> eZ A' of electrons scattered on nuclei due to the γ−A' mixing. The part of the primary beam energy is deposited in the ECAL1, while the rest of the total energy is transmitted by the A' through the "ECAL1 wall". The A' penetrates the ECAL1 and veto V1 without interactions and decays in flight in the decay volume DV into a narrow angle e+e− pair, which generates the second electromagnetic shower in the ECAL2 resulting in the two-shower signature in the detector. The sum of energies deposited in the ECAL1+ECAL2 is equal to the primary beam energy. The e+e- tracks are detected by two low-material budget straw tube stations ST1 and ST2. HCAL is the hadronic calorimeter.

The opening angle of the e+e- pair has the value $[\theta] \sim M(A')/E(A')$, where $M$ and $E$ are the mass and energy of the massive photon A'. Thus, if $M(A') \sim 50$ MeV and $E(A') = 30$ GeV, one obtains $[\theta] \sim 2$ mrad, and the distance between tracks of e+ and e- in the detector placed at $Lv = 5$ m downstream of the decay vertex is about $Lv[\theta] = 10$ mm. Under these conditions, the event produced in ECAL2 by e+e- pair cannot be resolved as a two-shower event and, therefore, will be registered as a single shower.

A feasibility study shows that detection of e+e- pairs and reconstruction of the decay vertex of the A' --> e+e- event in the vacuum decay volume (Fig. 1) would provide a large background suppression factor in the search for rare A'-> e+e- events. This makes the design and construction of a coordinate detector which can efficiently detect and resolve narrow e+e- pairs both in this and other experiments very interesting and important.

## 3. Coordinate detectors based on thin-wall drift tubes

### 3.1 Straw chambers for the P348 project

The proposed setup will use two stations of the coordinate detectors S1 and S2 based on straw tubes which have kapton walls of 70 µm or smaller. The first station, S1, will have X and Y two layer beam chambers with a sensitive area of 200x200 mm2 built of 6 mm diameter straws flushed with a gas mixture $ArCO_2$ (80/20) at a pressure of 1 bar. Maximum collection time of ionization electrons is about 60 ns. The chambers operate in the conventional for such detectors mode. They are wide spread and thus will not be discussed below.

The second station is intended for identification of e+e- pairs by registering separately the electron and positron in different straws. To this end, thin-wall tubes with an internal diameter of 2 mm have been manufactured by winding two strips of kapton film on a rod. Otherwise, they are identical to the tubes of the transition radiation detector of the ATLAS detector [4,5].

Figure 2 shows the parts of straw tubes of the inner diameter 4 mm and 2 mm which measured 520 mm in length, and a wall thickness of 72 µm (top), as well as the straw detector prototype made of 200 mm long and 2 mm diameter tubes (bottom). The parameters of the prototype have been investigated at a gas pressure varying from 1 bar to 3 bar. The anode wire is 20 µm gold plated tungsten. In the studies, we have used 5.9 keV gammas from a Fe-55 radioactive source and 3.55 MeV electrons from Ru-106 source.

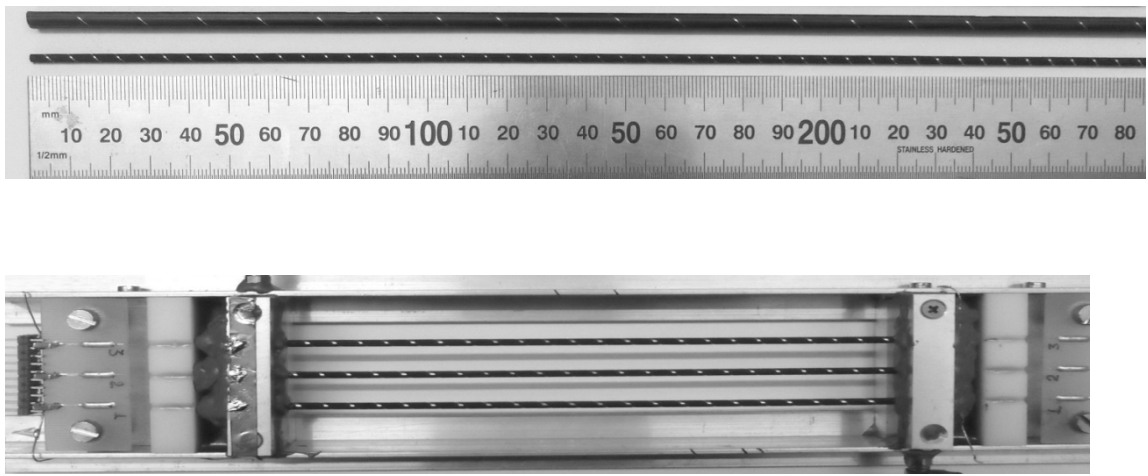

Fig.2. Thin-wall drift tubes of 4 and 2 mm diameter (top) and a prototype for the studies of the 2 mm straws (bottom).

The radiative thickness of the two layer drift chamber based on 2 mm straws (inner diameter) at a pressure of 3 bar and 6 mm straws at a pressure of 1 bar depending on the wall thickness of 72 μm and 40 μm is 0.109X0 and 0.056X0, respectively ( see Table 1).

**Table 1.** *Radiation thickness of a double-layer straw drift chamber, gas mixture $ArCO_2$ (80/20).*

| Diameter of straws, mm | P, bar | Material | Thickness, cm | Layers | $X_0$, cm | Radiation Thickness (%$X_0$) |
|---|---|---|---|---|---|---|
| 2 and 6 | | Kapton film | 0,0072 | 4 | 28,6 | 0,1007 |
| 2 and 6 | | | 0,0040 | | | 0,0559 |
| 2 and 6 | | Cu | $10^{-7}$ | 8 | 1,43 | $0,559 \times 10^{-4}$ |
| 2 and 6 | | W | $0,364 \times 10^{-4}$ | 2 | 0,35 | $0,897 \times 10^{-4}$ |
| 2 | 3 | Ar | 0,157 | 2 | 11000 | $68,51 \times 10^{-4}$ |
| 6 | 1 | | 0,471 | | | |
| 2 | 3 | $CO_2$ | 0,157 | 2 | 18300 | $10,29 \times 10^{-4}$ |
| 6 | 1 | | 0,471 | | | |

### 3.2. Parameters of the 2 mm straw tube detector

The parameters of the 2 mm straw tube detector have been compared to those obtained with the 4 mm straw detectors which reliably operate at the ATLAS TRT. The comparison has been performed by irradiating the detectors at a pressure of 1 bar in a wide beam of gammas from a Fe-55 source. The signal amplitude and energy resolution of the straw detector under study as a function of the anode voltage are shown in Fig. 3. The anode wire diameter in the straws of 4 and 2 mm was 30 and 20 μm, respectively. As is obvious, the amplitude dependencies are similar to each other, and the signal amplitude corresponding to the limited proportional mode in a 2 mm straw is even higher than that in a 4 mm straw by ~ 30%. A small deterioration of the energy resolution in a 2 mm straw can be tolerated. A deterioration of the energy resolution measured in a 4 mm straw at ~ 1600 V can be explained by the onset of a saturation of the detected signals when a Fe-55 source is used. The similarity in the measured parameters can serve as a justification of the program of constructing precision coordinate detectors based on straw tubes of small inner diameter.

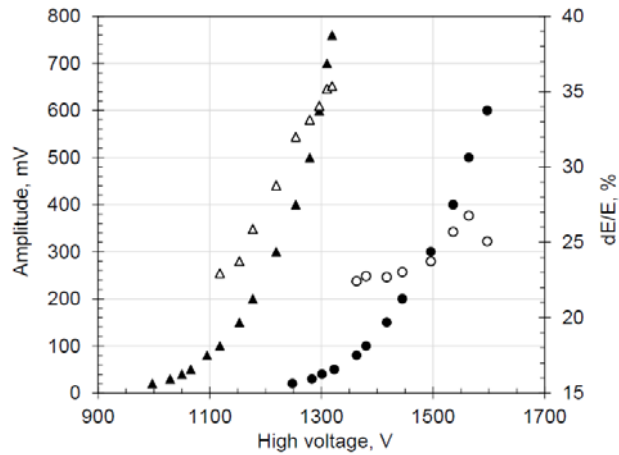

Fig. 3. Signal amplitudes (filled symbols) end the energy resolution (open symbols) as a function of the anode voltage for the straws of diameter 2 mm (triangles) and 4 mm (circles), respectively. Gamma quanta from the 55Fe source are registered in similar conditions by one amplifier; the gas mixture pressure is 1 bar.

Distributions of the ionization losses $\Delta E$ for 20 GeV electrons crossing one 2mm straw tube filled with the gas mixture in question at a pressure of 1 and 3 bar are shown in Fig. 4. The mean energy loss proved to be 0.6 keV and 1.6 keV for former and the latter case, respectively. It is evident that in some events the energy loss $\Delta E$ can be smaller than the detection threshold. The probability of the signal loss in such cases is shown in Fig. 5 as a function of the detection threshold. The data in Fig. 5 shown for different values of the gas mixture pressure indicate for rather high probability of the signal loss (particle detection inefficiency) at a pressure of 1 bar, whereas particles can be detected with efficiency of about 99% at a pressure of 3 bar.

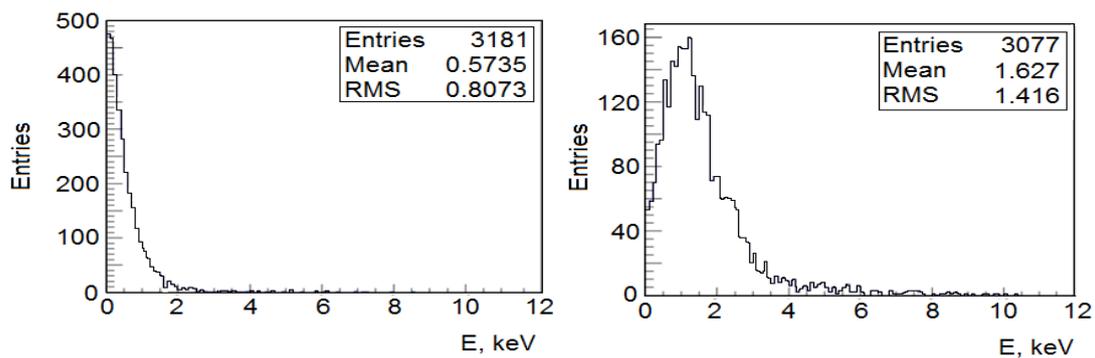

Fig.4. Distribution of the electron ionization losses in 2 mm straws at a gas pressure 1 bar (left panel) and 3 bar (right panel). A gas mixture ArCO2 (80/20) is used.

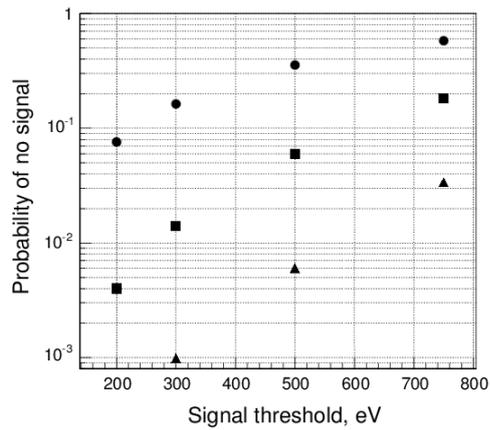

Fig.5. The probability of a signal loss as a function of the detection threshold in the straw tubes which further will be related to thresholds of the registering electronics. Gas mixture pressure is 1 (●), 2 (■) and 3 (▲) bar.

It should be underlined that the detection efficiency for minimum ionizing particles (MIP) proved to be different for straw tubes of different diameter. As has been measured in the tests of straw detectors conducted at the CERN SPS, the efficiency for a gas mixture ArCO2 (80/20) was 99% and 87% for the straws of diameter 9.53 mm and 4 mm, respectively [6,7]. The efficiency of the ATLAS TRT straw detector for the similar gas mixture is ~ 80% [8].

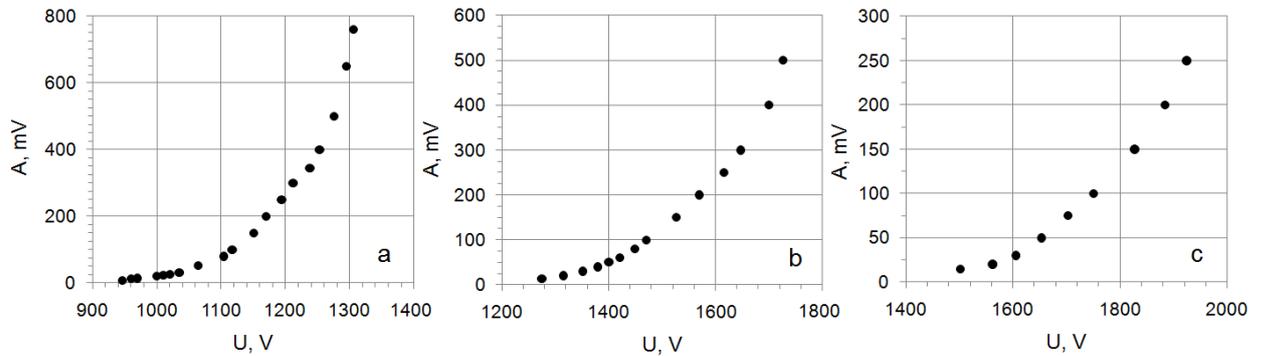

Fig, 6. Signal amplitude from the 2 mm straw for gas mixture ArCO2 (80/20) obtained by registering 5.9 keV gamma quanta. Left, central, and right panels correspond to the gas mixture pressure of 1bar, 2 bar, and 3 bar, respectively.

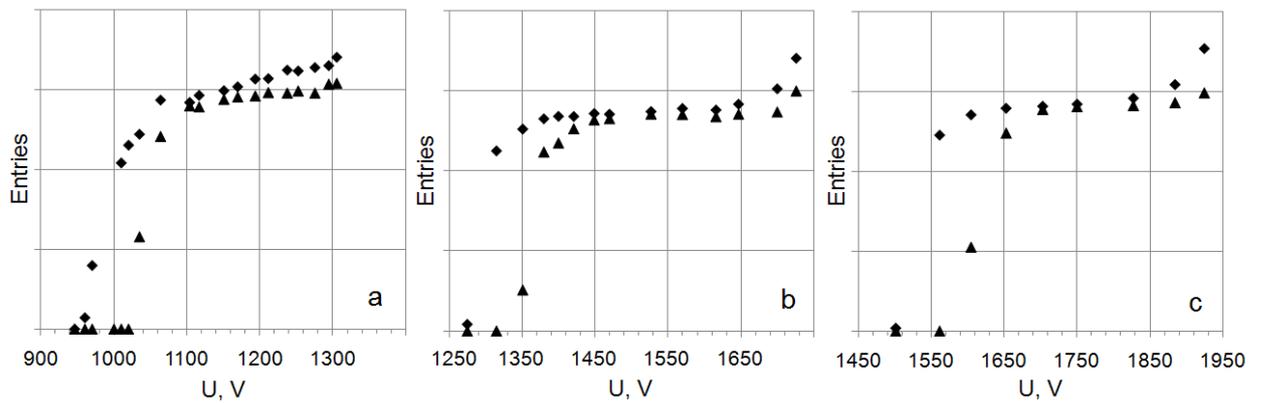

Fig.7. Counting rate as a function of the high voltage for the 2 mm straw tubes filled with ArCO2 (80/20) gas mixture obtained by registering 5.9 keV gamma quanta. Left, central, and right panels correspond to the gas mixture pressure of 1bar, 2 bar, and 3 bar, respectively. The signal detection thresholds were 12 mV (diamonds) and 26 mV (triangles).

The signal amplitude and counting rate as a function of the anode voltage for 2 mm straw tubes studied at three different gas pressure are displayed in Figs. 6 and 7, respectively. The amplitude dependence presented in Fig. 6 corresponds to the proportional and limited proportional gas amplification mode. Signal amplitudes as high as ~ 700, 500 and 300 mV are detected at a gas pressure of 1, 2, and 3 bar, respectively. The count rate plateau estimated from the results displayed in Fig. 7 is larger than 200 V if there are no background counts.

The detection efficiency for 3.55 MeV electrons as a function of a gas pressure has been measured by using electrons emitted by the Ru-106 source. It is displayed in relative units in Fig. 8. In all measurements, the anode voltage corresponded to the center of the count rate plateau (Fig. 7). It is evident that the efficiency drops by ~ 10% and ~ 60% when the gas pressure decreases from 3 bar to 2 and 1 bar, respectively.

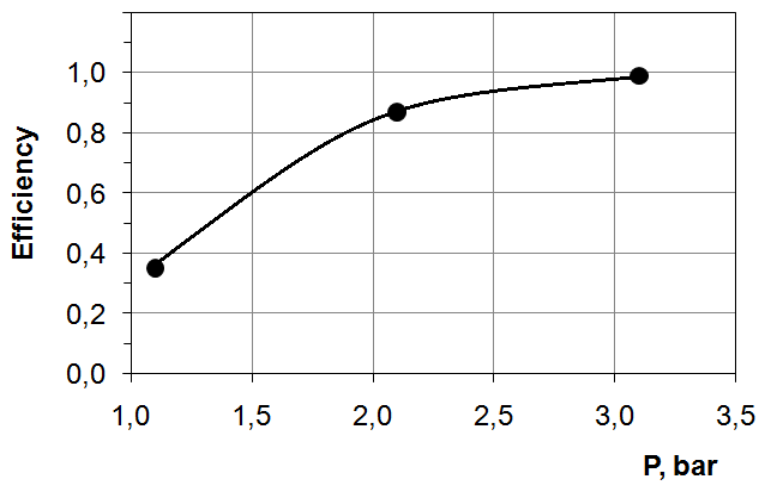

Fig.8. Detection efficiency for 3.55 MeV electrons registered in 2 mm straw tubes as a function of the pressure of the ArCO2 (80/20) gas mixture.

## 3.3 Spatial resolution in the detection of electron-positron pairs with small opening angle

Due to small angular divergence of electron-positron pairs produced in the decay volume of the experimental setup, the coordinate detectors should provide a good two-hit resolution. The particle which passes through a two-layer chamber (see Fig. 9, left panel), crosses the straw tubes in each of two layers, and the sum of the ionization electron drift times measured in the tubes is equal to the maximum electron collection time.

If a straw is crossed by two particles, two points in which the electromagnetic wave is generated are produced at the anode wire with the time interval of Delta t = (~ 20 x Delta r) ns, where Delta r is the distance between the particle track and the anode wire in units of mm, and 20 ns is the time needed for ionized electrons to pass a 1 mm distance in a gas mixture based on argon at a pressure of 1 bar. A superposition of the two waves generates a signal which does not generally contain information on the number of passing particles. The contribution of propagation time of the electromagnetic waves along the anode to the superimposed signal is very negligible due to short time of passage which is 35 ps for 1 cm distance [9].

If a chamber is traversed by several particles separated by a distance larger than the straw diameter, 2$r$, they can be detected independently by different straw tubes. In the event of passage of two particles separated by a distance larger than $r$, they can be registered either in one layer by two adjacent straws or/and by the straws of the first and second layers shifted by $r$.

Possible events of passage of a pair of charged particles separated by a distance smaller than $r$ through a two-layer detector are shown schematically in Fig. 9. It is possible that a pair of particles passes between the anodes of two closest straw tubes of the first, L1, and second, L2, layer (Z-1 zone). It is also possible that the two particles of the same pair pass closer than at the distance of $r/2$ to the anode wire of the straw belonging to the first or second layer (Z-2 zone). It is clear from Fig. 9 that in the former case the first particle of the pair is detected by the straw of one layer, and the second one is registered in the straw of another layer. In the latter case, the closest to the anode particles are detected by the closest straws from different layers (particles 1 and 2) or by two adjacent straws in the same layer (pairs of particles 1 – 2 and 2 – 3). Therefore, the use of the straws of diameter 2 mm makes it possible to construct detectors which possess two-hit resolution of the order of 0.5 mm.

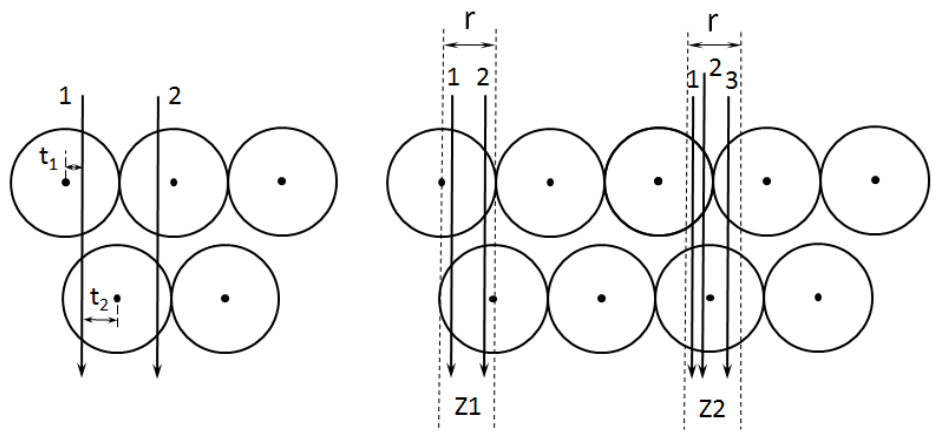

Fig.9. Schematic view of the options for which simultaneous detection of two close particles is possible.

The minimal achieved magnitude of the resolution of two-hit events is determined by the maximum collection time of ionization electrons $T$ and the spatial resolution of the straw detector.

## 4. Conclusions

Occupancy is an important detector parameter. It is possible to decrease the straw detector occupancy by decreasing the drift tube diameters, however, not dramatically. The use of large diameter straw tubes with segmented anodes (granulated straw) can decrease the counting rate by a substantially larger factor and to effectively detect particles at a gas pressure of 1 bar [7,10]. Therefore, new capabilities of the coordinate detectors constructed based on the thin-wall drift tubes of 2 mm inner diameter and operated at a gas filling pressure of 2 – 3 bar basically allow one to

- provide the high detection efficiency of the minimum ionizing particles;
- provide the two-hit resolution distance not worse than 0.5 mm;
- increase the intrinsic spatial resolution of the straws;
- decrease the straw wall thickness down to 30 µm by using kapton films of 12 or 7 µm.

**References**

[1]. J.L. Hewett et al., arXiv:1401.6077 [hep-ph]

[2]. S.N. Gninenko, Phys. Rev. D 89 (2014) 7, 075008; S. Andreas et al. arXive: 1312.3309 [hep-ex]; CERN-SPSC-2013-034 / SPSC-P348.

[3]. * L. Shehtman. Private communication.

[4]. T. P. A. Akesson, et al. (RD6 Collab.), "The ATLAS TRT Straw Proportional Tubes: Performance at Very High Counting Rate," Nucl.Instr. Met. A., V. 367. P143-153 (1995).

[5]. Yu. V. Gusakov, N. Grigalashvili, F. Dittus, G. D. Kekelidze, V. M. Lysan, V. V. Myalkovskii, V. D. Peshekhonov, N. A. Rusakovich, A. A. Savenkov, D. Froidevaux, and E. M. Khabarova, "Atlas TRT — Research & design B-type module mass production," Phys. Part. Nucl. **41**, 1–26 (2010).

[6]. V.I. Davkov, I.Gregor, D.Haas, S.V.Mouraviev, V.V.Myalkovskiy , L.Naumann, V.D. Peshekhonov, C.Rembser, I.A.Rufanov, N.A.Russakovich, P.Senger, S.Yu. Smirnov, V.O.Tikhomirov. "Spatial resolution of thin-walled high-pressure drift tubes", Nucl.Instrum. Methods Phys. Res., A 634, 5–7 (2011).

[7]. S. N. Bazylev, K. I. Davkov, I. Gregor, D. Haas, S. V. Mouraviev, V. V. Myalkovskiy, L. Naumann, V. D. Peshekhonov, C. Rembser, I. A. Rufanov, N. A. Russakovich, P. Senger, A. V.

Shutov, I. V. Slepnev, S. Yu. Smirnov, V. O. Tikhomirov, and I. A. Zhukov, "A prototype coordinate detector based on granulated thin_walled drift tubes," Nucl. Instrum. Methods Phys.Res., A 632, 75–80 (2011).

[8]. E. Abat, T. N. Addy, T. P. A. Akesson, et al. (ATLAS TRT Collab.), "The ATLAS Transition RadiationTracker (TRT) proportional drift tube: design and performance," J. Instrum. **3**, P02013 (2008).

[9]. V. D. Peshekhonov. "Coordinate Detectors Based on Thin_Wall Drift Tubes". *Physics of Particles and Nuclei, 2015, Vol. 46, No. 1, pp. 94–122.*

[10]. K. Davkov, V. Davkov, R. Geyer, Y.V. Gusakov, G.D. Kekelidze, V.V. Myalkovskiy, L. Naumann, D.V. Peshekhonov, V.D. Peshekhonov, A.A. Savenkov, V.O. Tikhomirov, K.S. Viryasov. "Development of segmented straws for very high-rate capability coordinate detector" Nucl. Instrum. and Methods Phys. Res., A 584, 285–290 (2008).